\documentclass[final,5p,times,twocolumn]{elsarticle}

\usepackage{graphicx}
\usepackage{aas_macros}
\usepackage{xspace} 

\usepackage{amssymb}
\usepackage[utf8]{inputenc}
\usepackage[T1]{fontenc}

\usepackage{tikz}
\usetikzlibrary{shapes,arrows,chains}

\usepackage[ruled,vlined]{algorithm2e}

\newcommand{\F}{{\mathcal F}}
\newcommand{\Fo}{{\mathcal F}_0}
\newcommand{\sky}{{\ttfamily PolGrawAllSky}\xspace}
\newcommand{\skyfarm}{{\ttfamily skyfarmer}\xspace}

\journal{Computer Physics Communications}

\begin{document}

\begin{frontmatter}



\title{Architecture, implementation and parallelization of the software to search for periodic gravitational wave signals}

\author[scc]{G.~Poghosyan\corref{corrauth}}
\ead{Gevorg.Poghosyan@kit.edu}
\cortext[corrauth]{Corresponding author: Tel:+4972160825604 Fax:+4972160824972 }
\author[scc,pec]{S.~Matta}
\author[scc]{A.~Streit}

\address[scc]{Steinbuch Centre for Computing, Karlsruhe Institute of Technology, 76131  Karlsruhe, Germany}
\address[pec]{PEC University of Technology, 160012 Chandigarh, India}

\author[camk]{M.~Bejger}
\address[camk]{N. Copernicus Astronomical Center of the Polish Academy of Sciences, 00-716 Warsaw, Poland}

\author[im]{A.~Kr\'olak}
\address[im]{Institute of Mathematics of the Polish Academy of Sciences, 00-956 Warsaw, Poland}

\begin{abstract}
The parallelization, design and scalability of the \sky code to search for periodic gravitational waves from rotating neutron stars is discussed. 
The code is based on an efficient implementation of the $\F$-statistic using the Fast Fourier Transform algorithm.
To perform an analysis of data from the advanced LIGO and Virgo gravitational wave detectors' network, which will start operating in 2015, hundreds of millions of CPU hours will be required - the code utilizing the potential of massively parallel supercomputers is therefore mandatory.
We have parallelized the code using the Message Passing Interface standard, implemented a mechanism for combining the searches at different sky-positions and frequency bands into one extremely scalable program. 
The parallel I/O interface is used to escape bottlenecks, when writing the generated data into file system. 
This allowed to develop a highly scalable computation code, which would enable the data analysis at large scales on acceptable time scales. 
Benchmarking of the code on a Cray XE6 system was performed to show efficiency of our parallelization concept and to demonstrate scaling up to 50 thousand cores in parallel.
\end{abstract}


\end{frontmatter}

{\bf PROGRAM SUMMARY}

\begin{small}
\noindent
{\em Manuscript Title:} Architecture, implementation and parallelization of the software to search for periodic gravitational wave signals\\
{\em Authors:} G.Poghosyan, S.Matta, A.Streit, M.Bejger, A.Kr\'olak     \\
{\em Program Title:} parallel \sky                                      \\
{\em Journal Reference:}                                                \\
{\em Catalogue identifier:}                                             \\
{\em Licensing provisions:} none                                        \\
{\em Programming language:} C                                           \\
{\em Computer:} Any parallel computing platform supporting MPI standard \\
{\em Operating system:} Linux as well any other supporting MPI standard \\
{\em RAM:} 1 Gigabyte per parallel task                                 \\
{\em Number of processors used:} tested with up to 50 208 processors              
{\em Classification:} 1.5 Relativity and Gravitation                     \\
{\em External routines/libraries:} MPI v.2 or newer, FFTW v.3 or newer  \\
{\em Nature of problem:} Search for periodic gravitational waves from rotating neutron stars     \\
{\em Solution method:} The $\F$-statistic method using the Fast Fourier Transform algorithm    \\
   \\

\end{small}

\section{Introduction}
\label{sect:into}

Gravitational waves (GWs) - variations of the curvature of spacetime, able to
propagate through spacetime in a wave-like fashion - were first predicted by
Albert Einstein \cite{Ein1916}, and they are a direct consequence of the
general theory of relativity that he proposed. Several properties of GWs are
similar to those of electromagnetic waves. GWs also propagate with speed of
light and are polarized (two polarizations in the description of general relativity). The best empirical, yet 
{\it indirect} evidence for gravitational radiation comes form the
observations of tight relativistic binary pulsar systems; first such a system
was discovered by R. Hulse and J. Taylor with the radio observations 
from the Arecibo telescope \cite{HT75}. 
Direct detection of GWs will constitute a very
precise test of Einstein's theory of relativity and open a new field - GW
astronomy. Currently, the most promising GW detector concept is of the
Michelson-Morley interferometer type. The detection principle is as follows: 
while a GW passes through such a detector, it changes the length of its arms 
and affects the interference pattern of the laser light circulating in 
the interferometer \cite{Saul94}.

State-of-the-art interferometric GW detectors,
LIGO\footnote{http://www.ligo.org} in the USA and Europe
(Italian-French, with the contribution of Hungary, the Netherlands and
Poland) Virgo\footnote{https://wwwcascina.virgo.infn.it} have collected a
large amount of data that are still being analyzed. Meanwhile, the advanced
LIGO and Virgo detectors are under construction and they are forecasted to
start collecting new, more sensitive data at the end of 2015. It is expected
that these advanced detectors will be sufficiently sensitive so that the
direct detection of GWs can finally be achieved. As the GW signals are extremely
weak, their detection constitutes a major challenge in data analysis and
computing. Several types of astrophysical GW sources are investigated: 
coalescence of compact binaries containing neutron stars and black holes, 
supernova explosions, quantum effects in the early Universe as well as 
rotating, non-axisymmetric neutron stars. 

The departure from axisymmetry in the mass distribution of a rotating neutron
star can be caused by strong magnetic fields and/or elastic
stresses in its interior. The search for such long-lived, periodic 
GW signals generated by the spinning star is nevertheless 
particularly computationally intensive. 
This is because the GW signal is very weak and one needs to analyze 
long stretches of data in order to extract the signal ''buried'' in 
the noise of the detector. Due to this, the modulation
of the signal due to the motion of the detector with respect to the solar
system barycenter has to be taken into account; it depends on the location of
the source and a modulation that is a function of the intrinsic change of
rotation frequency of the deformed neutron star. Moreover, we do not know the
polarization, amplitude, and phase of the GW signal. Consequently, the
parameter space to search for the signal becomes very large.

The {\em Polgraw-Virgo} team, working within the LIGO scientific
Collaboration (LSC) and the Virgo Collaboration has developed algorithms and
a pipeline called \sky to search for GW signals from spinning
neutron stars \cite{Astone2010}. The pipeline was applied to the analysis of
the data gathered by the Virgo detector during its first science run denoted
as VSR1. The analysis involved 5 million CPU hours and took almost three
years to complete \cite{Aasi:2014}. The serial code's design allowed to use only
one processor core and was run on a number of computer clusters with standard
queuing systems. This performance turned out to be not entirely satisfactory
for current and future requirements of the GW data analysis. To analyze all
the data collected by the Virgo detector, 250 million CPU hours are required, 
whereas the analysis of all the data that will be collected 
from the advanced detectors expected to be available by the year 2018 
will require four times more resources, i.e., 1000 million CPU hours.
To perform this analysis one would need 1petaFLOPS computer 
working continuously for one year.

\begin{figure}[ht]
	\centering
	\includegraphics[width=\columnwidth]{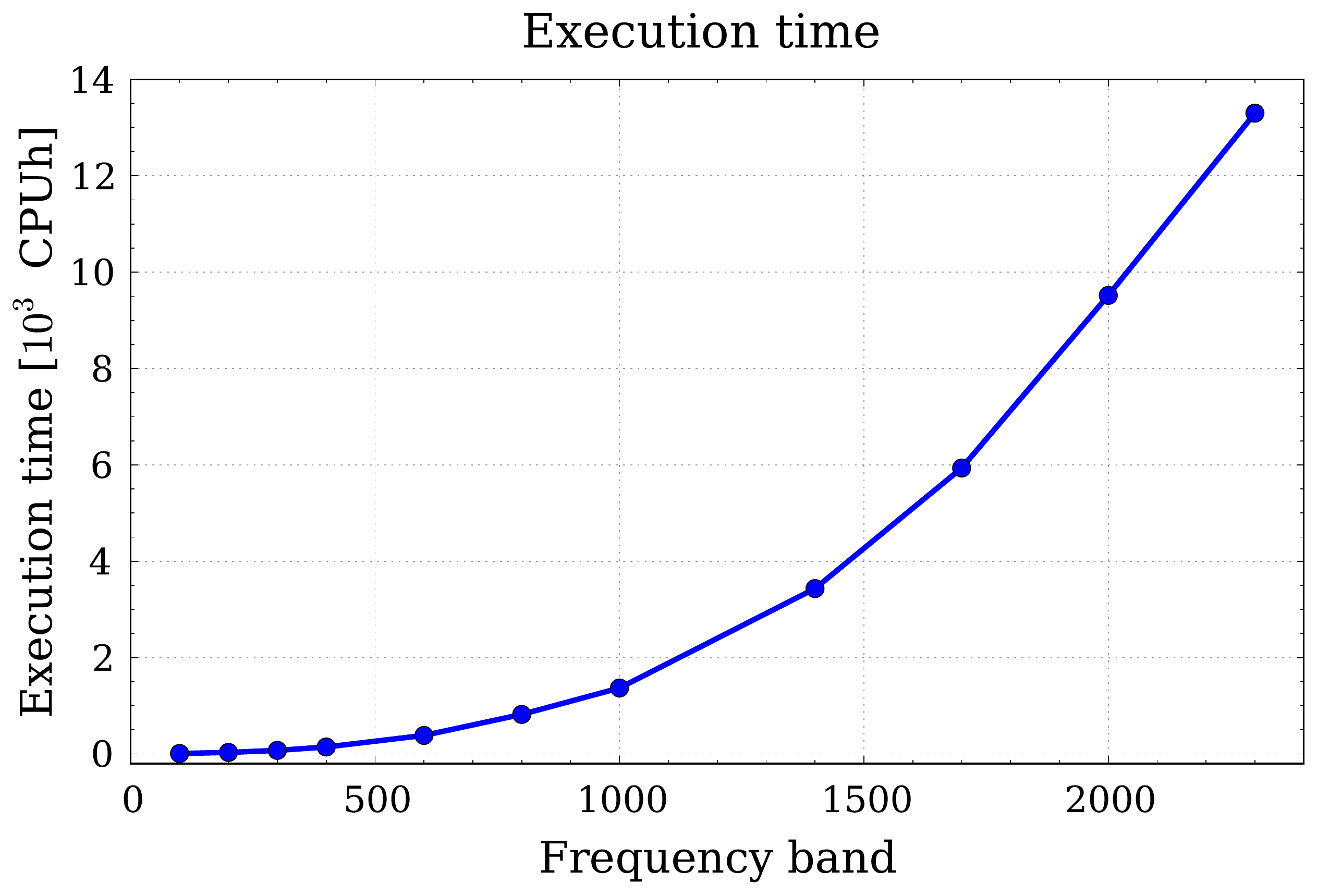}
	\caption{Total execution time of embedded \sky code in thousand CPU hours as a function of frequency band.}
	\label{fig:exeband}
\end{figure}

\begin{figure}[t]
	\centering
	\includegraphics[width=\columnwidth]{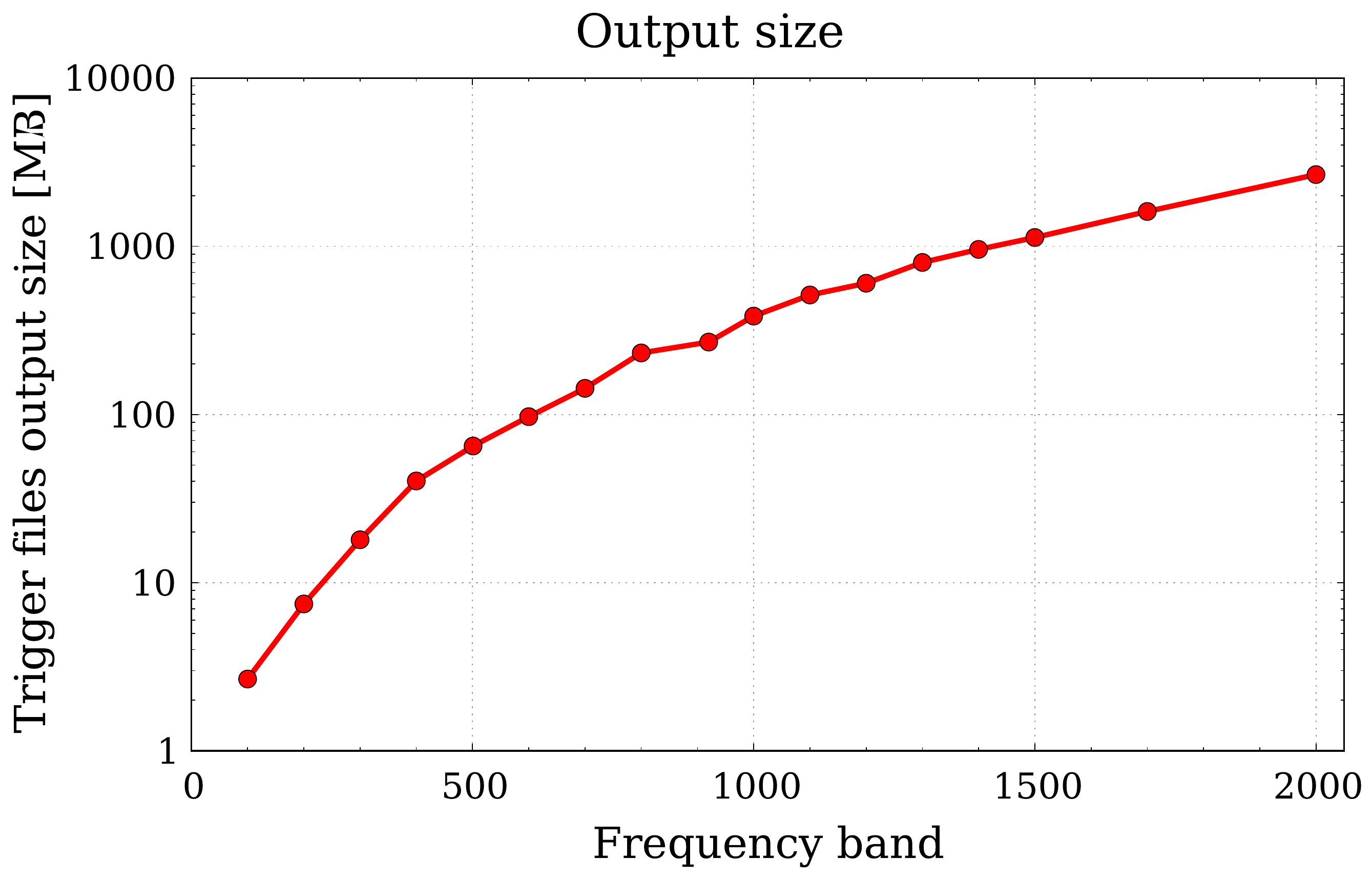}
	\caption{Total output of \sky code  in Megabytes as a function of frequency band.}
	\label{fig:outband}
\end{figure}

To estimate the computational requirements we have performed
representative tests with the Gaussian noise data at different band
frequencies illustrated on Fig.~\ref{fig:exeband} and \ref{fig:outband}.
For example, a serial search for GWs at frequencies 600, 1000, 1700 and 2000
will require a total of 20 thousand CPU hours of computation, which is
more than two years on a single CPU and correspondingly the output
generated by this simulation would be ca. 4GB.

To alleviate the computations, this paper proposes a massively 
parallelized version of the \sky
code that uses the Message Passing Interface (MPI) library
\cite{MPISpec3}. 
MPI is a distributed memory parallelization scheme commonly used in high
performance computing (HPC). This parallelized version  is able to run on high
performance computers with tens of thousands of cores. We are reporting 
a sufficient performance increase of the parallel \sky code enabling its
usage on massively parallel HPC systems for production analysis of data already
collected by GW detectors and also of data from the advanced GW detectors that
will start to be available by the end of the year 2015 \cite{Aasi:2013a}.

\subsection{Mathematical Methodology}
\label{ssect:fstat}

The algorithms to search for gravitational wave signals from rotating neutron
stars implement the $\F$-statistic \cite{JKS1998}, derived by one of us and
commonly adopted in other pipelines (see e.g., \cite{AasiEHS52013}). 
By using the $\F$-statistic one doesn't need to search for the
polarization, amplitude and phase of the signal. Instead, one is left 
with a 4-dimensional space parameterized by the GW frequency, frequency 
derivative (spindown, reflecting the fact that the pulsar is spinning down) 
and the two angles determining the location of the source in the sky. 
To implement a computationally efficient algorithm we are faced with 
two problems. Firstly one would like to minimize the number of grid
points on which the $\F$-statistic is evaluated, achieving at the same 
time a certain target sensitivity of the search. 
This is equivalent to a well-known geometrical problem called 
the {\it covering problem}. Secondly, we would like to take advantage 
of the Fast Fourier Transform (FFT) algorithm.
The FFT algorithm cannot be directly implemented in the calculation of the
$\F$-statistic because of the modulation of the signal due to the
detector's movement around the Sun. In order to implement
it one needs to interpolate the data (so called re-sampling). However,
the re-sampling means an additional computational cost that can offset the 
advantage of the FFT algorithm. Moreover, the FFT algorithm evaluates 
the $\F$-statistic at some specific values of the frequency called the 
Fourier frequencies. These frequencies need to be reconciled with the 
grid points obtained by the solution of the covering problem. 
\cite{Astone2010} describes in detail how the covering problem 
and efficient usage of the FFT can be achieved: by constraining the
solution it is ensured that one needs to perform the computationally intensive
re-sampling procedure only once per sky position. Each sky position 
corresponds to a grid of values of frequency derivative - the re-sampling 
occupies therefore only a fraction of the total computational
time. Thus we have an algorithm that involves both the evaluation 
of the $\F$-statistic at smallest number of points and takes the 
advantage of the FFT at the same time. 

The main computation consists of 3 loops - two external loops over the 
sky positions, an inner loop over the frequency derivatives (spindowns) 
and finally the FFT execution that evaluates the $\F$-statistic on a grid 
of frequencies.

\section{Description of the code for searching of GW signals}
\label{sect:corecode}

The data time series from a detector is divided into two-day time slots, 
numbered by an integer $d$, that we call {\em frames}.  
A typical number of time frames in a science run is around one hundred. 
Each frame of data is divided into narrow frequency bands of 1 Hz width 
each. The bands overlap by $2^{-5}$~Hz and they are numbered by integer
$b$. The relation between the frequency $f_{\rm off}$ of the lower edge 
of the band and the band number $b$ is given by
\begin{equation}
f_{\rm off} = 100 + (1 - 2^{-5}) b.
\end{equation}
Given that the interferometric detectors span the frequency range
between $\sim\,10$ and $\sim\,1000$ Hz, in a typical search the number 
of bands is around one thousand.  Thus one has about 
{\it one hundred thousand} of time-frequency data sets to analyze.

The \sky code analyzes a given band $b$ in a given time frame $d$. The input
data consist of three files: narrowband time series of the detector data {\tt
xdat\_d\_b.bin}, ephemeris of the detector {\tt DetSSB.bin} and grid generating
matrix {\tt grid.bin}.  The data {\tt xdat\_d\_b.bin} spans the length of 2
sidereal days and is sampled at $0.5$ s, thus consisting of $N = 344656$
double precision numbers. 

Detector ephemeris {\tt DetSSB.bin} file contains the 3-dimensional
vector, relating the detector to the Solar System Barycenter (SSB), of
the same length as the time series data, as well as two additional
parameters: phase $\phi_o$ determining the position of the detector at
the time of the first sample of the {\tt xdat\_d\_b.bin} file, and
$\epsilon$, which is the obliquity of the ecliptic.  Thus {\tt
DetSSB.bin} contains $3 \times N + 2 = 1033970$ double precision
numbers. 

The file {\tt grid.bin} contains a $4 \times 4$ grid lattice generating
matrix $M$. The rows of the matrix $M$ are four vectors spanning our
4-dimensional parameter space. The integer multiples of these vectors
define the grid points in the parameter space where the $\F$-statistic
is calculated. As already mentioned in Sect.~\ref{ssect:fstat}, the code
performs 3 loops - two external ones over the sky positions, and an
inner loop over the range of frequency derivatives (spindowns). The sky
positions are transformed from the usual astronomical equatorial
coordinates into two integers ${\bf n}$ and ${\bf m}$ and every spindown
value is transformed to an integer denoted by ${\bf s}$. 

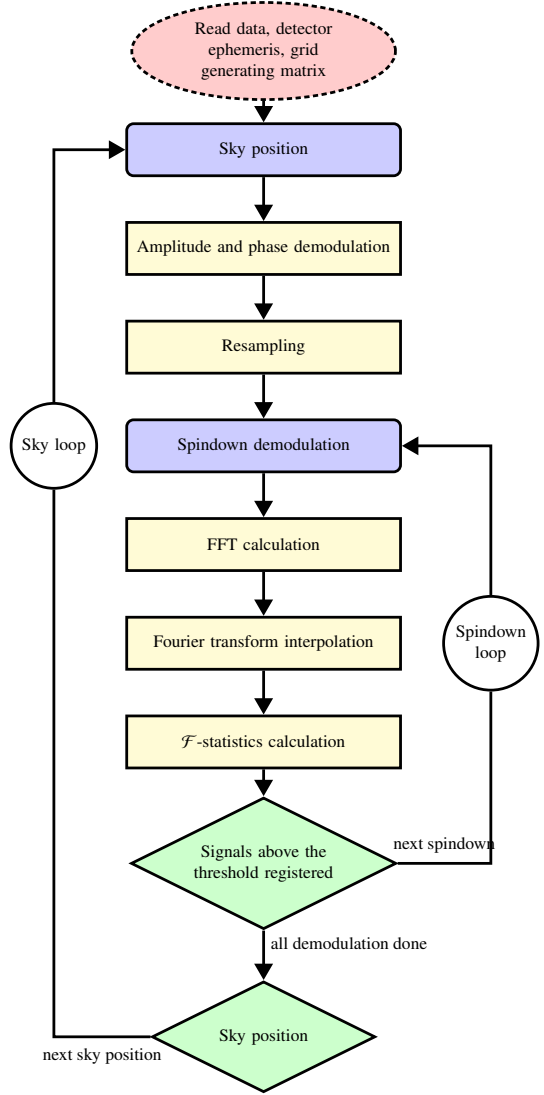
\begin{figure}[t]
	\centering
\tikzset{
base/.style={draw, ultra thick, on grid, align=center, minimum height=3em, text width=15em},
brick/.style = {base, fill=yellow!20},
block/.style = {brick, fill=blue!20, rounded corners},
decision/.style = {base, fill=green!20, diamond, text width=7em},
line/.style = {->, draw, ultra thick, >=triangle 60}
}
\resizebox{0.8\columnwidth} {!} {
\begin{tikzpicture}[node distance = 2cm, minimum height=2em, auto]
    \node [ellipse, densely dashed, base, fill=red!20, text width=10em] (init)
        {Read data, detector ephemeris, grid generating matrix};
    \node [block, below of=init] (sky) {Sky position};
	    \path [line] (init) -- (sky);
    \node [brick, below of=sky] (skydemodul) {Amplitude and phase demodulation};
	    \path [line] (sky) -- (skydemodul);
	\node [brick, below of=skydemodul] (resample) {Resampling};
	    \path [line] (skydemodul) -- (resample);    
    \node [block, below of=resample] (demodul) {Spindown demodulation};
	    \path [line] (resample) -- (demodul);
		    \node [circle, base , left of=demodul, node distance=12em, text width=4em ] (skyloop) {Sky loop};
	\node [brick, below of=demodul] (fft) {FFT calculation};
	    \path [line] (demodul) -- (fft);	    
    \node [brick, below of=fft] (FFT) {Fourier transform interpolation};
	    \path [line] (fft) -- (FFT);
		    \node [circle, base , right of=FFT, node distance=13em, text width=4em ] (spin) {Spindown loop};
    \node [brick, below of=FFT] (Fstat) {$\mathcal{F}$-statistics calculation};
	    \path [line] (FFT) --  (Fstat);
    \node [decision, aspect=2, below of=Fstat, node distance=7em, text width=9em] (decide) {Signals above the threshold registered};
	    \path [line] (Fstat) -- (decide);
   	 	\path [line] (decide.east) -| node [near start] {next spindown} (spin) |- (demodul.east);
    \node [decision, aspect=2, below of=decide, node distance=10em, text width=9em] (nextsky) {Sky position};
	    \path [line] (nextsky.west) -| node [near start] {next sky position} (skyloop) |- (sky.west);
	    \path [line] (decide) -- node [near start] {all demodulation done} (nextsky);
\end{tikzpicture}
}

        \caption{Flow diagram of the core code for searching GW signals.}
        \label{fig:coreflowdiag}
\end{figure}

The flow diagram of the main \sky code, underlying its most important
features, is presented in Fig.~\ref{fig:coreflowdiag}; a detailed
description can be found in \cite{Astone2010}. The first step is to
read the data and store it in the memory. Then we start the two loops
over the sky, i.e., the range of integers ${\bf n}, {\bf m}$ described
by the grid.  First the sky position in equatorial coordinates is
recovered. Then the two amplitude modulations and the phase modulation
are calculated and the signal is demodulated. Then we perform resampling
(interpolations) using the FFT and splines. The search grid is
constructed in such a way that we need to perform resampling only once
for each sky position, thus allowing for re-using the same resampled and
demodulated data in the inner spindown loop. For each ${\bf s}$ value in
the inner spindown loop we perform the demodulation.  This is followed
by two Fourier transforms, each for one amplitude demodulation. We then
perform interpolation of the FFTs resulting in Fourier transform twice
as long as the original one. This interpolation is applied in order to
prevent excessive loss of the signal-to-noise ratio as the parameters of
the true signal do not necessarily coincide with parameters of the grid
points (\cite{Astone2010}, Sect. VIB).  Finally, for each ${\bf s}$
value the $\F$-statistic is calculated. Whenever the value of the
$\F$-statistic crosses a predetermined threshold $\Fo$, we register the
parameters of this grid point (sky position, frequency and spindown),
together with the value of the $\F$-statistic.  This set of 5 double
precision numbers constitutes the {\it candidate signal} output. The
candidate signals obtained from the analysis of {\tt xdat\_d\_b.bin}
data are stored in the file named {\tt candidates\_d\_b.bin}. The
candidate files are then subject of analysis by post-processing codes to
extract true gravitational wave signals (if no statistically significant
gravitational wave signal is found, an upper limits for the amplitude of
the gravitational wave at a given frequency can be obtained).

\section{Parallelization of sky loops}
\label{sect:para}

As the sky positions are independent of each other, this feature can be
exploited to parallelize the outer (sky) loop (see
Fig.~\ref{fig:coreflowdiag}).  Current parallel version keeps the inner
(spindown) loop over the frequency derivatives. We recall that the inner
loop reuses the demodulated and re-sampled data, as described in
\cite{Astone2010}. Our choice of the parameter space is such that the
number of frequency spindowns is a linear function of the band frequency
$f$, whereas the total number of sky positions $N_{\rm sky}$ is a quadratic
function of the band frequency. It is evident that the relation 
$N_{\rm sky}(f)$ plays a crucial role in the parallelization strategy 
of our computations.

\begin{figure}[ht]
	\centering
	\includegraphics[width=\columnwidth]{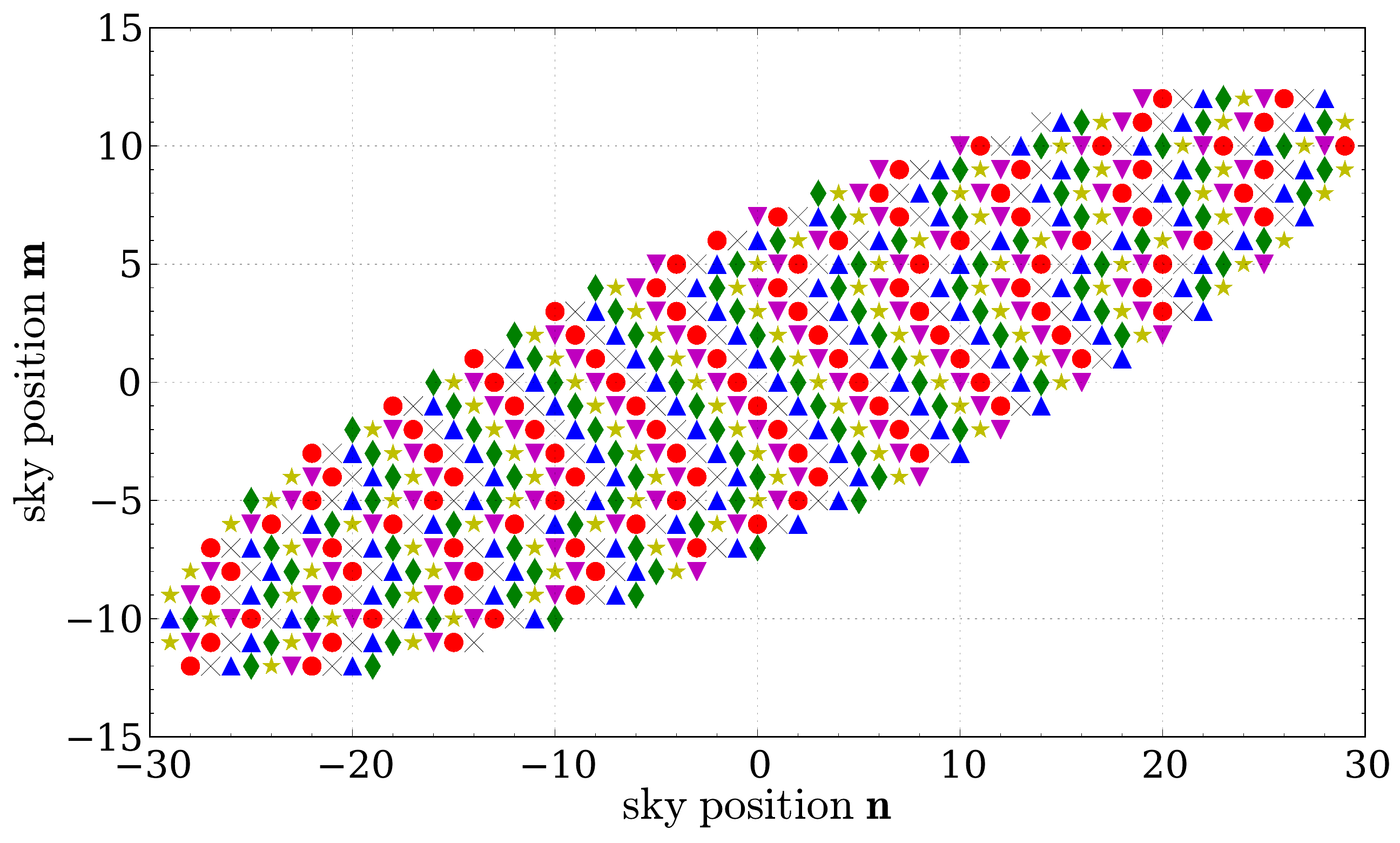}
	\caption{Distribution of computations of the sky positions for frequency band $b = 100$ and
	for $N_{tasks} = 6$. Six parallel tasks are represented by different symbols and are repeatedly covering the whole       
	hemisphere based on the {\tt round-robin} scheduling algorithm.}
	\label{fig:parsky}
\end{figure}

Computation of spindowns for each sky position is distributed onto the
available parallel tasks using the MPI point-to-point and collective
communication routines implemented in it. Each parallel task makes
$P_{\rm size}$ sky positions computations, where $P_{\rm size} = \lceil
N_{\rm sky}/N_{\rm tasks} \rceil$ is the ratio of the number of sky
positions to the number of parallel tasks, rounded up to the nearest 
integer. To reach optimal load balancing, computations are distributed 
to parallel tasks (see Fig.~\ref{fig:parsky}) using the {\tt round-robin}
(RR) scheduling algorithm \cite{RoundRobin}.

The tests of scaling performance of the parallel \sky code on computer
clusters with up to 1000 CPU cores were encouraging, but not entirely
satisfactory. Such facilities are nowadays available in many universities
or research computing centers. However, on small size supercomputers one needs to compute 
continuously without maintenance for several years in order to analyze
a significant fraction of data collected by the LIGO and Virgo
detectors. Hence, the only solution is to increase the number of
parallel tasks by at least an order of magnitude to be able to speak
about acceptable execution times for the full analysis of the data.

\begin{figure}[ht]
	\centering
	\includegraphics[width=\columnwidth]{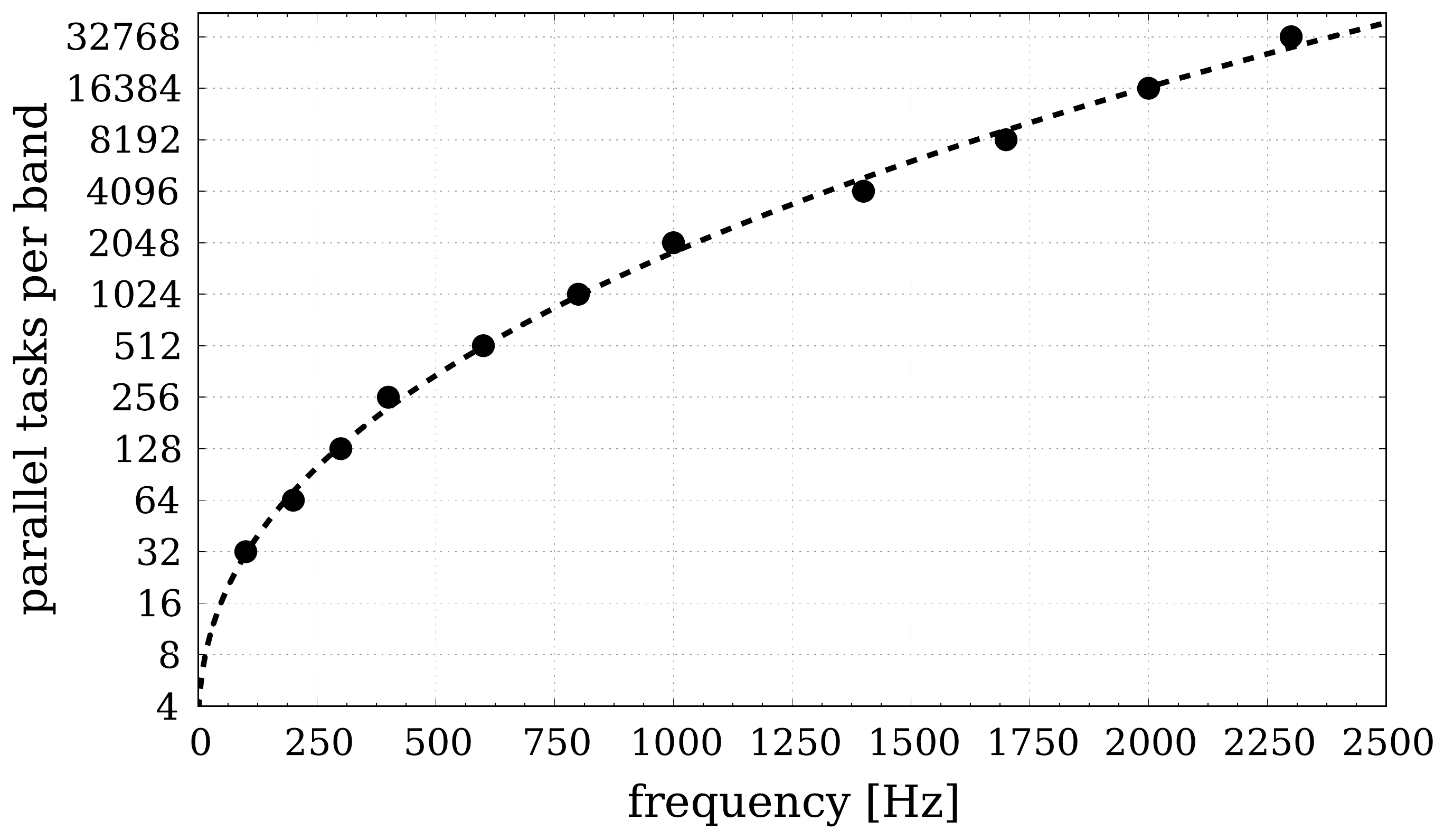}
	\caption{Optimal amount of parallel tasks for a given frequency band. The number of parallel tasks
	given on the y-axis is required in order to analyze the corresponding frequency band on the x-axis within an hour.}
	\label{fig:cpuband}
\end{figure}

Using a limited amount of available computational resources provided
within the framework of a PRACE project\footnote{PRACE Preparatory
Access type B 2010PA1183 for 50 000 core-hours on GCS@HLRS, Germany} we
have tested the scalability of the parallelized \sky code for frequencies 
up to 2300 Hz, using up to 32768 parallel tasks. The PRACE grant 
was also useful to estimate the optimal number of parallel tasks 
per frequency band, when the time necessary to complete 
the analysis of any band was restricted (an example for one hour
restriction time is shown in Fig.~\ref{fig:cpuband}).

Estimation of the optimal amount of parallel tasks per frequency
band allowed us to combine a set of parallel computations of sky loops
into one, in order to perform the computation on massively parallel processing
({\tt MPP}) systems.  We have used the fact that not only the sky positions, but 
also searches for GW signal candidates at different frequency bands are
independent of each other. Such a low coupled system allows the use of 
distributed network computing systems such as volunteer or
grid/cloud computing. Contrary to distributed systems, executing 
parallel tasks on {\tt MPP} systems make scheduling and work-flow
problems easier to avoid, e.g., by using built-in schedulers 
and process manager mechanisms specific for the {\tt MPP} systems \cite{dynmpi}.

\subsection{Algorithmic outline of \skyfarm}
\label{sect:paralgo}
To enhance the scalability of execution of many computations in
parallel, we combine many instances consisting of different \sky
executions that use different numbers of parallel sub-tasks. 
This feature is implemented using the dynamic process creation 
and grouping framework of MPI, with 
different MPI sub-worlds also known as virtual groups, that 
enables collective and dynamic communication operations across 
a subset of related tasks. 
The main \sky code with parallel sky loop is encapsulated into
another code, named \skyfarm, equipped with internal scheduling 
and bookkeeping mechanism. The \skyfarm flow chart is depicted in 
Fig.~\ref{fig:skyfarm}. 
For each sky search a domain group and associated communicator
as a new virtual MPI sub-world is created, which allows 
to execute instances of \sky without global source re-engineering.  
This concept, known as algorithmic skeleton (AS), was proposed 
for parallel programming to simplify not only the programming but
also to enhance the portability of parallel applications by abstracting 
from the underlying hardware \cite{colemit,Cole2004389}. We have
developed our own version of farm skeleton \skyfarm that is optimized 
for running the \sky and is able to reach higher scalability 
on multi-node system using MPI, as compared to other AS tools 
\cite{Ernsting2012129,Gonzalez-Velez20101135}.
The new communicators are simply created and managed by \skyfarm 
transparently from the point of view of the embedded parallel \sky, 
whose tasks run as if in a stand-alone mode with a uniquely 
assigned communicator.  

\pgfdeclarelayer{marx}
\pgfsetlayers{main,marx}
\providecommand{\cmark}[2][]{%
  \begin{pgfonlayer}{marx}
    \node [nmark] at (c#2#1) {#2};
  \end{pgfonlayer}{marx}
  }
\providecommand{\cmark}[2][]{\relax}
\begin{figure*}
	\centering
\tikzset{
base/.style={draw, on grid, align=center, minimum height=3em, text width=16em , minimum height=2em, node distance=6em, ultra thick},
brick/.style = {base, fill=red!20},
block/.style = {brick, fill=blue!20, rounded corners},
dis/.style = {diamond, base, fill=green!20, text width=7em, aspect=3, node distance=8em},
elli/.style = {ellipse, base, densely dashed, fill=red!20},
cir/.style = {circle, base, densely dashed, fill=blue!20},
line/.style = {->, draw, ultra thick, >=triangle 60}
}
\resizebox {0.75\textwidth} {!} {
\begin{tikzpicture}[auto]
	
    \node [brick] (init) {Initialize MPI environment and read input parameters};
    \node [brick, right of=init, xshift=15em] (starter) {Identify type of each parallel task};
	    \path [line] (init.east) -- (starter.west);
    \node [dis, below of=starter] (idcheck) {Check own type};
	    \path [line] (starter) -- node [left] {Test messaging all-to-all} (idcheck);
   	\node [brick, left of=idcheck, xshift=-15.5em] (grupi) {Identify group size per frequency band};
	    \path [line] (idcheck) -- node [xshift=1em] {{\tt MASTER}} (grupi);
   	\node [brick, below of=grupi] (distrib) {Distribute bands to groups};
	    \path [line] (grupi) -- node [near start, left] {Initialize bookkeeping} (distrib);
   	\node [dis, below of=distrib, text width=8em] (schedul) {Schedule bands to {\tt SLAVE}s};
	    \path [line] (distrib) -- (schedul);
		\node [elli, right of=schedul, text width=8em, node distance=21em] (allfree) {Send group size and band value to all free {\tt SLAVE}s};
			\path [line] (schedul.east) -- node [text width=4em, yshift=-2.5em] {Bands in queue {\tt SLAVEs} free} (allfree);
			\path [line] (allfree.north) |- node [near end, text width=14em, yshift=1.5em, xshift=2.5em] {Back to bookkeeping} (distrib);
		\node [block, below of=allfree, text width=14em, node distance=10em] (freeslave) {Wait for message from {\tt MASTER} };
				\path [line] (idcheck.east) --++(8mm,-55mm) node [left,near end] {\tt SLAVE} -- (freeslave.north);
				\path [line,dashdotted, ultra thick] (allfree.south) -| node [near end, left, text width=5em, yshift=0.5em] {Message including a TAG} (freeslave.north);
			\node [dis, below of=freeslave, yshift=2.7em] (tagcheck) {Check the TAG};
				\path [line] (freeslave) -- (tagcheck);
			\node [brick, below of=tagcheck, text width=10em, node distance=6em] (initgroup) {Initialize subgroup and own communicator};
				\path [line] (tagcheck) -- node [near end, right, text width=7em, yshift=0.7em] {TAG $\neq$ "DIE"} (initgroup);
			\node [elli, below of=initgroup, text width=7em, node distance=5em] (jobNallnsky) {Parallelized \sky};
				\path [line, ultra thick] (initgroup) -- (jobNallnsky);
				\path [line] (jobNallnsky.east) --++(13mm,0mm) |- (freeslave.east);
	\node [block, below of=schedul, text width=14em, node distance=10em] (wait4free) {Wait for message from {\tt SLAVE}};
		\path [line] (schedul) -- node [text width=4em] {Bands still in queue} node [left, text width=4em] {{\tt SLAVEs} still busy}(wait4free);
		\path [line,dashdotted, ultra thick] (jobNallnsky.west) -- (wait4free.east);
   	\node [brick, below of=wait4free, node distance=8em] (storenew) {Set {\tt SLAVE} state free in bookkeeping};
	    \path [line] (wait4free) -- node [near start, right, text width=10em, yshift=-2em] {{\tt SLAVE} finished a band} (storenew);
		\path [line] (storenew.west)  --++ (-5mm,0mm) |- (distrib);
    \node [elli, text width=8em, below of=storenew] (finish) {Send "DIE" TAG to {\tt SLAVE}s};
	    \path [line, dashdotted, ultra thick] (finish.east) --++(21mm,0mm) node [near end, xshift=-2em] {TAG="DIE"} |- (freeslave.west);
        \path [line] (schedul.west)  --++(-21mm,0mm)  node [near start, above, text width=7em] {Band needs more {\tt SLAVE}s than available} node [near start, below, text width=7em] {Queue empty and {\tt SLAVE}s free} |- (finish.west);
    \node [brick, text width=8em, below of=finish] (final) {Finalize};
	    \path [line]  (finish.south) -| (final);
	    \path [line] (tagcheck.west) |- node [above,xshift=-3em] {TAG="DIE"} (final.east);	
\end{tikzpicture}
}

\caption{Flow diagram of the \skyfarm for running \sky at 
different frequency bands in separate MPI worlds/groups. Bricks
represent usual algorithmic steps; diamonds are steps with decisions
about future direction of run; ellipses are steps when message passing
take place represented as dash-dotted paths; blocks with rounded corners
represent steps in which {\tt MASTER} or {\tt SLAVE} ranks are waiting
until a new MPI message is received.}
        \label{fig:skyfarm}

\end{figure*}
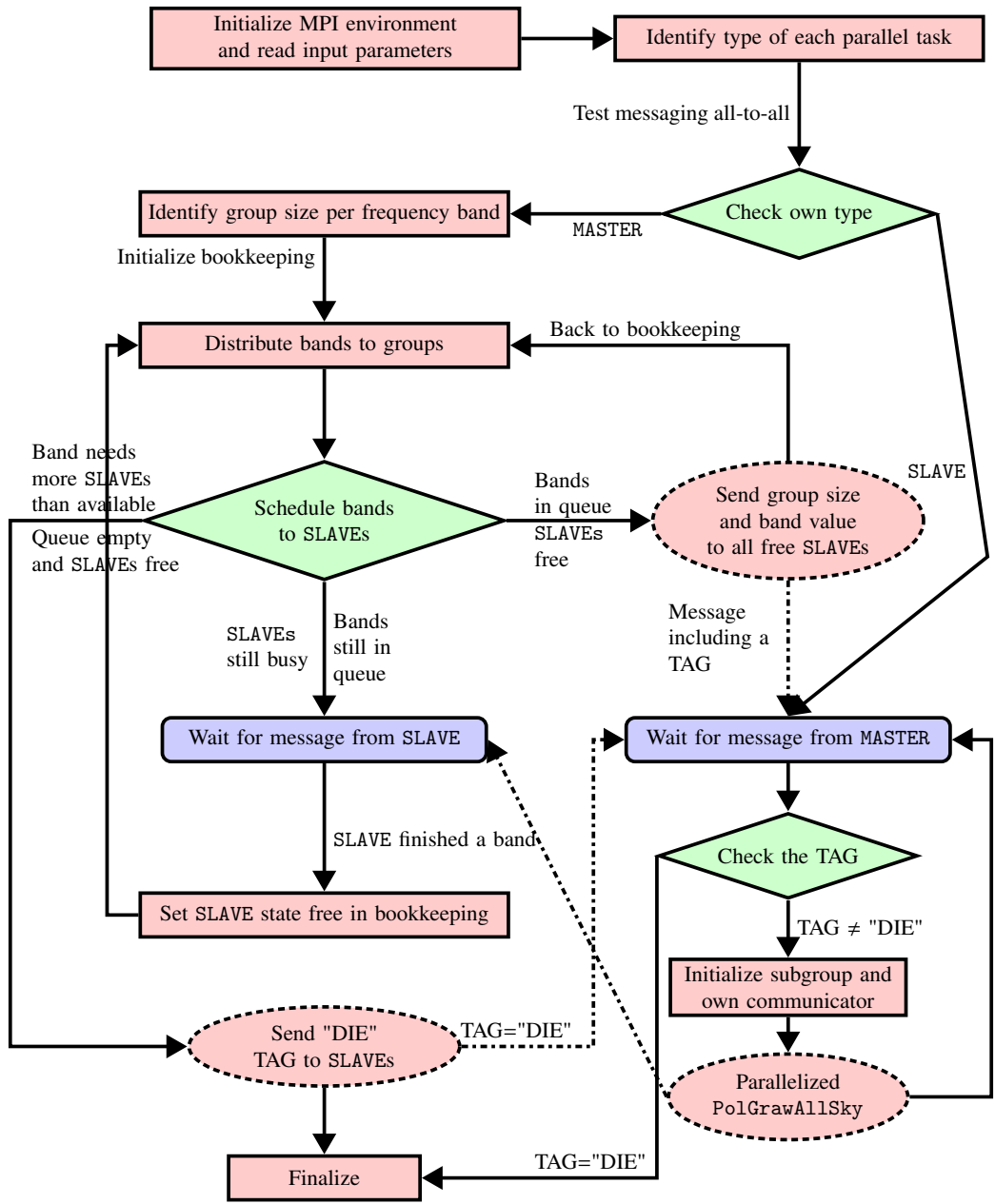

The structure of the \skyfarm is divided into five main parts:
\begin{enumerate}
\item initialization and estimation of the available 
	and necessary parallel resources,
\item construction of different tasks as groups,
\item distribution and decomposition of groups,
\item bookkeeping information about free and busy resources,
\item execution of the \sky code.
\end{enumerate}
After the MPI environment is successfully initialized, the virtual
grouping execution model is carried on and the master-slave model
becomes the basic architecture of the simulation. 
The computing resources are split into non-equal parallel groups
to ensure that the execution of all embedded \sky codes takes
approximately the same time, in order to achieve proper load balance.
The implementation for global information exchange for scheduling and bookkeeping is
based on MPI all-to-all non-blocking communication.

\subsubsection{Domain decomposition}
{\tt MPP} with internal scheduling and work-flow may be prone to 
problems of load balancing, especially when 
many parallel computations are embedded into one big run. 
This typically results in limiting scalability. 
Therefore, to optimally use big {\tt MPP} systems with more than
ten thousand parallel tasks, we have implemented a domain decomposition
algorithm based on estimation of optimal number of parallel tasks per
frequency band.

The execution of \skyfarm starts with reading various parameters like
frame number, input data directory, output data directory and two input
files {\em cpuperband.dat} containing two columns - frequency $f_k$ and
the corresponding estimation for optimal number of parallel tasks
$S(f_k)$, as well the file {\em frequency.dat} containing the list of frequency
bands to be analyzed (see the Algorithm \ref{algomaster}). These files are
used to compute group size $S(f)$ for a given frequency band $f$ and
generate a set of encapsulated parallel runs. The size of group
of tasks to be used by various bands is estimated using first
order spline-polynomial interpolation
\begin{equation}
\label{eq:spline}
S(f) = a \cdot f + b \\
  = \underbrace{\frac{S_2 - S_1}{f_2 - f_1}}_{a} \cdot f
     + \underbrace{S_1 - \frac{S_2 - S_1}{f_2 - f_1} \cdot f_1}_{b \,=\, S_{{}_1}-a\cdot f_{{}_1}},
\end{equation}
where $f_1$ and $f_2$ are two frequencies in the {\em cpuperband.dat}
file neighboring $f$ and $S(f_1)$ and $S(f_2)$ are the corresponding 
numbers of parallel tasks.
In our tests we used an estimation presented in Fig.~\ref{fig:cpuband}
and generated on the basis of scalability tests on a computing
cluster with up to 50 000 computing cores with 10 GigaFLOPS performance each.
This estimation may of course be changed or recomputed if the load balance 
for each frequency band needs to be different than one hour. For instance, 
the values for the $S(f_k)$ per band could be reduced to perform the computation
longer than in one hour, when the performance computing cores is weaker
or simply to fit to amount of available parallel cores of particular MPP systems.

{\small
\begin{algorithm}[h]
\label{algomaster}
 \caption{Algorithm for initialization, domain decomposition and bookkeeping}
	\SetAlgoLined
	\KwData{{\tt MASTER} rank}
	\KwResult{Preparation of parallel simulations}
	Read and sort the band table\;
	Initializing the bookkeeping\;
	Define the domains as groups\;
	\While{infinitely}
	{
	Check and count free slaves\;
	\If{not enough free slaves}{go into waiting state}
	Check bookkeeping for bands to run\;
	\eIf{all groups fit}
		{
		\For{each group}
			{
			Send group size and band value to slaves\;
			Create new communicators\;
			Set the slaves as busy\;
			Set the groups as submitted\;
			}
		}
		{\eIf{biggest group does not fit}
			{Stop simulation\;}
			{\eIf{{\tt SLAVE}s are busy}
				{
				Go into waiting state\;
				Finished message received\;
				Set slaves free\;
				}
				{Finish simulation}
			}
		}
    }
\end{algorithm}
}

This procedure aids to estimate the resources - number of tasks to be divided into virtual groups for the multi-level parallel runs of \skyfarm and parallelized \sky codes encapsulated into \skyfarm.
If any of the frequency bands listed in file {\em frequency.dat} would require all available resources, then only this band will
run. 
The corresponding virtual group will occupy all the resources and the search of GW signals at other frequency bands will not be performed in parallel.
This results to single level parallelism and could be counted as an almost sequential run even if each search at a given frequency would be still running in parallel.

In the current version of the \skyfarm a case of global insufficiency
appears and no simulation for frequencies given in {\em frequency.dat}
will be carried out when the number of available resources is less than
is necessary for the highest frequency case. So the estimate based on
{\em cpuperband.dat} must allow to compose virtual groups for each
frequency given in {\em frequency.dat}, even if this group will occupy
the whole available resources. 

\subsubsection{Bookkeeping algorithm - scheduling for multi-level
parallelization}
For advanced scalability one always needs to provide more 
resources than the summary of all the tasks for each frequency search.
In this way all embedded \sky codes may run in parallel and optimal
speed-up, i.e. the reduction of the {\em wall time} of execution will be
reached.  Alas, it is not always possible to provide as much resources as 
it is needed to complete all computation in parallel at once, therefore some
sub-sequential runs of groups are organized by the scheduling and
bookkeeping mechanism implemented in \skyfarm.

The {\tt MASTER} process initializes a bookkeeping algorithm that 
keeps the information about the slave tasks' status and tracks 
the subgroups in queue, allowing them to run as soon as enough 
resources are available. During a run the {\tt MASTER} stores 
the status of {\tt SLAVE} processes as {\em busy} or {\em free} 
in a {\em book} array, the index of which corresponds to initial 
rank of task. At start all are initialized as free except the 
{\tt MASTER} task, which is always marked busy. 
Subsequently, the {\tt MASTER} distributes the information about the defined
groups with corresponding band frequency to {\tt SLAVEs}, switches their
corresponding states in bookkeeping array into busy and enters into infinite
waiting state expecting messages from any {\tt SLAVE}.  As soon as a message
tag "FINISHED" is received from {\tt SLAVEs} of a given group, the status of all ranks in
group, whom the slave belongs to are changed to free. Following, a check of the
status of all the {\tt SLAVEs} maintained in the structure book is made, and as
a result the additional {\tt free\_ranks} array is populated with the ranks of
slaves found in bookkeeping array stated as free. The size of the {\tt
free\_ranks} array is used to identify computation for any other frequency band waiting
in the queue and with a group size that may fit the available free resources.

However, if meanwhile some resources were freed, 
the creation of a new communicator for a new group to start a new parallel simulation
in addition to the already running ones is not possible due to use of
collective mechanism for communicator creation in current version of
\skyfarm - one needs to send the information about a new group
redistribution of tasks to all tasks and this is possible only when all
tasks are in the free state. In order to be able to create 
a new virtual group we have used {\tt MPI\_Comm\_create}, 
which should be called simultaneously by all MPI ranks belonging 
to parent communicator. 

As a result, one has to check at every new iteration if all the ranks
and groups have finished their partial simulation and are in status free,
before starting any new set of parallel groups. This limitation
results in a loss of scalability. The waiting time for other
unfinished parallel simulations (even if there are enough free resources
available for a new simulation) may be decreased by using the {\em
intercommunicators} \cite{dyngrp}. Herewith the communicator creation of
virtual groups is collective over those processes only that will be
members in the resulting communicator. But disadvantage of the non-collective
communicator creation is the {\it group creation cost} - time spent 
to create a new group increases exponentially depending on the number 
of parallel processes, because of the recursive merging nature of the 
algorithm. This could be amortized however by a potential benefit 
to particular application and hence for case of \skyfarm should be extra investigated.

\subsubsection{Algorithm for execution of embedded code in parallel}
Like the {\tt MASTER}, the {\tt SLAVEs} are placed in an infinite loop which
breaks only when the whole simulation is finished, i.e., when it 
reaches the box {\tt Finalize} in Fig.~\ref{fig:skyfarm}. 
Initially, the {\tt SLAVEs} are in the waiting state. 
When the first receive statement comes, it contains not only 
the particular information about groups to run, but also a tag. 
In the case of tag is "DIE", the infinite loop will be
broken, otherwise it means there is a group, in which the {\tt SLAVE} 
must participate in a joint simulation. After receiving the membership of 
a respective group, all the {\tt SLAVEs} place the communicator
creation call with respect of their group; if the {\tt SLAVE} does not
find itself as a member of any group, then the group creation call
returns {\tt MPI\_COMM\_NULL}, and such {\tt SLAVE} is being marked 
by {\tt MASTER} in bookkeeping as a "FREE". 

When the {\tt SLAVE} identifies itself as a member of one of the groups,
then it receives the respective frequency
band, passed further to the embedded \sky code.  After
completing the execution the group frees their communicator and 
a "FREE" status tag is sent to the {\tt MASTER} from all 
the {\tt SLAVEs} members of the group. The cycle continues as long as 
all the bands are completed and the "DIE" tag is received by 
the {\tt MASTER}. This is summarized by the Algorithm \ref{algoslave}.

{\small
\begin{algorithm}[h]
\label{algoslave}
 \caption{Algorithm for running parallel embedded codes}
	\SetAlgoLined
	\KwData{{\tt SLAVE} rank}
	\KwResult{Generation of parallel simulations}
	\While{infinitely}
	{
	Receive information about all groups and status tag\;
	\eIf{status tag is not DIE}
	{
		\For{check all group}
			{
			Receive band frequency\;
			Receive particular group size\;
			Receive ranks of group\;
			Initialize and try to create group communicator\;
			\If{New communicator is one whom slave must belong to}
			{Save the group size, band and communicator information}
			}
		\If{New communicator is found to whom {\tt SLAVE}  must belong to}
		{
			Call the embedded code using identified communicator, band frequency\;
			Free particular group and communicator used from \sky code
		}
		Send FREE status to {\tt MASTER}
	}
	{
		Finish calculation
	}
    }
\end{algorithm}
}

\subsection{Implementing parallel Input/Output interface}
At large parallelization scale, the limitations of codes based on trivial farm
skeleton paradigms are inevitable, leading to the loss of performance.  It is 
usually due to the Input/Output ({\tt I/O}) activity rising up as
''bottlenecks''. The execution of embedded code on {\tt MPP} systems with 
more than 1000 processing units faces such limitations of parallel file 
systems. Even if the encapsulated code is rarely using {\tt I/O}, 
the performance of the farm is limited directly by the 
scalability of the file system. The fact that each parallel-running 
embedded code writes its results as separate files results in intensive
usage of data storage. The final results are distributed in as many separate 
files as there are parallel tasks in farm. Handling of large number of files is then an additional 
problem in any MPP and moreover distributed computing systems.

To eliminate this problem we have implemented a parallel
writing mechanism to join the outputs (files with candidate signals). 
The parameters of signal candidates
are initially stored in the local memory of parallel {\tt SLAVEs} 
running different sky loops and at the end of computation the data are 
written in a single file per frequency band and frame, 
using the concept of collective operations supported by MPI {\tt I/O}. This 
functionality is implemented at various levels of the parallel code by using 
joint MPI files that are created at the start of each computation. 
Each parallel process obtains its part of the file for writing in the data 
of the candidate signals found in its analysis.

{\tt I/O} algorithm starts with the acquisition of its own view of joint MPI file by
each process participating in the operation. A view is defined in terms of three 
parameters: a displacement or location in the file given by the number of bytes
from the beginning of the file, an elementary data type and a file type. Before
the given process begins writing into joint MPI file in parallel, an 
{\tt MPI\_Allgather} collective communication is used to count signals that 
every {\tt SLAVE} has computed and stored in its local memory. This allows 
each {\tt SLAVE} to calculate the offset from which the {\tt SLAVE} starts 
writing. The offset is obtained by counting the number of 5 doubles that 
every other process with lower rank has to write for each identified signal.

\section{Performance and scalability analysis}
\label{sect:perf}
Productive usage of any parallel code on a supercomputing system with
thousands of tasks triggers a multitude of challenges significantly different
from those which rise when running the same code in a sequential way, even if
the amount of sequential executions is the same. 
When comparing a high throughput computation with sequential runs, the
benefit of using parallelized code on massively parallel systems is the
performance achieved by smaller time necessary to accomplish each parallel run.
It may however be limited due to bottlenecks or inherently sequential parts of
the parallelized application (Amdahl's Law, \cite{Rodgers}).

Unavoidably, the present version of parallelized code for GW signal searches  
has a maximal number of tasks up to which a speed-up can
be gained. For instance, even essentially eliminating non-scaling elements in the 
embedded \sky code was not enough to reach high efficiency at tests with more than thousand
parallel tasks. At higher scales the loss of performance due to the
{\tt I/O} activities was inevitable independently on using multi-site or single-site 
parallel computing systems. Non-optimal {\tt I/O} activities  
strongly limits the scalability of any code. To overcome this one has to implement a
system for storing the data in memory or writing them out at some point of
computation using the {\tt I/O} libraries optimized for parallel file systems. 
The corresponding {\tt I/O} implementation for the \sky code was 
described in the previous section.

Even after the elimination of the bottlenecks, the scalability of the parallelized \sky is limited due to the simplicity of the algorithm. 
Particularly, low coupling of partial analysis performed in the \sky code, allowed us  to keep interprocess communication limited and reach higher 
speed-up with more than 30 000 cores for most complex and computationally-intensive cases at frequencies as high as 2300 Hz. 
But depending on parameter space and the search frequency, there is a maximum scalability at which the \sky code can be used in parallel (Fig.~\ref{fig:cpuband}).

Hitherto, to extend the performance and scalability of the whole computation to levels above 30 thousand parallel tasks, it is crucial to increase the complexity of problem by combining many runs of the \sky code into one job.
By embedding the parallelized \sky into \skyfarm and implementing the
domain decomposition and efficient bookkeeping algorithm described in section
\ref{sect:paralgo}, we were able to reach scalability as high as $50 176$
parallel tasks for a test run when searching GW signal for 11 different
frequencies [100, 200, 300, 400, 600, 800, 1000, 1400, 1700, 2000, 2300] in one massive
parallel run.

To estimate efficiency of the parallelized \sky as well as of the \skyfarm code we have
made a performance analysis and scalability tests using the CRAY XE6
supercomputer at the High Performance Computing Center Stuttgart (HLRS), called 
Hermit \footnote{https://www.hlrs.de/systems/platforms/cray-xe6-hermit}. This
cluster consists of 3552 nodes interconnected through an InfiniBand network
with 113 664 compute cores altogether and 1 petaFLOPS peak performance.

\subsection {Strong and weak scaling}
\begin{figure}
	\centering
	\includegraphics[width=\columnwidth]{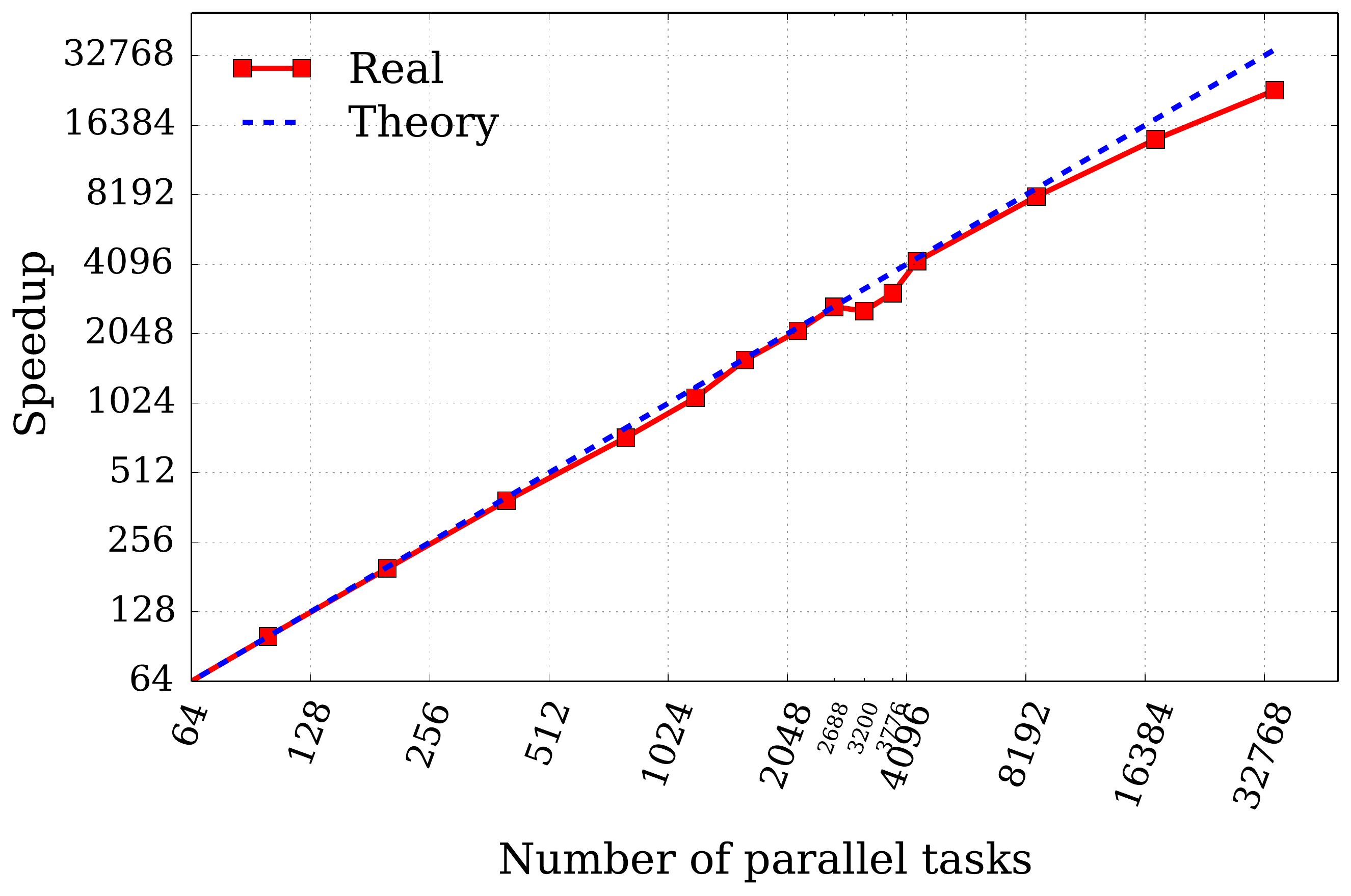}
	\caption{Strong scaling of the \skyfarm for joint computation of 7 
	frequency bands at once, using different amount of parallel tasks.}
	\label{fig:strskyfarm}
\end{figure}

By running a fixed-size problem on a varying number of processors one can see
how the timing of the computation scales with the number of processors and estimate 
what is the part of the code that is efficiently parallelized. This feature is
called the {\em strong scaling}. Fig.~\ref{fig:cpuband} shows the maximal scalability
of the embedded \sky code based on {\em strong scaling} tests performed for each given frequency band. 
Speed-up tests: reduction of the number of CPU hours
for the same-size simulations by increasing the number of parallel tasks, 
performed with the \skyfarm are presented in Fig.~\ref{fig:strskyfarm}.

In each run we have 
analyzed 7 frequency bands ranging from 100 to 700 Hz, with a 100 Hz step. 
The results show that our current implementation that uses MPI is able to 
run with up to 32 000 tasks in parallel very efficiently. 

However, as the number of computations of spindowns for each sky position per
parallel task decreases, the scaling is detached from the ideal linear speed-up,
because the communication per task starts to dominate the computation time. 
Also, not optimal algorithms of the scheduling and domain decomposition of
current version of \skyfarm degrade the performance. In other words 
the problem is not ''heavy'' enough to keep bigger amount of parallel 
processes ''busy'' with computation, as well as the mechanism of 
the group size estimation per chosen
frequency for running parallel MPI worlds, described in section
\ref{sect:paralgo}, is not fully optimal yet.

Secondly, the available tasks were not always perfectly fitting 
into the number of parallel tasks necessary for running a given 7 frequency 
bands as the domain size per frequency is estimated by first order 
spline-polynomial interpolation (see Eq.\ref{eq:spline}) and must be 
rounded down to the nearest integer. As a result, some tasks were simply 
not used. This is visible in Fig.~\ref{fig:strskyfarm} when the speed-up ''flattens out'' 
at around 4000 tasks, but picks up again when available tasks are optimally used.

The strong scaling test shows the advantages and disadvantages of using bookkeeping 
and the domain decomposition mechanism. On one hand they are enabling the large 
computation even if not enough resources are provided to analyze 
all frequencies in parallel at once, but on the other hand the optimal speed-up is 
reached only if the optimization mechanisms balance the needed and available 
resources.

\begin{figure}
	\centering
	\includegraphics[width=\columnwidth]{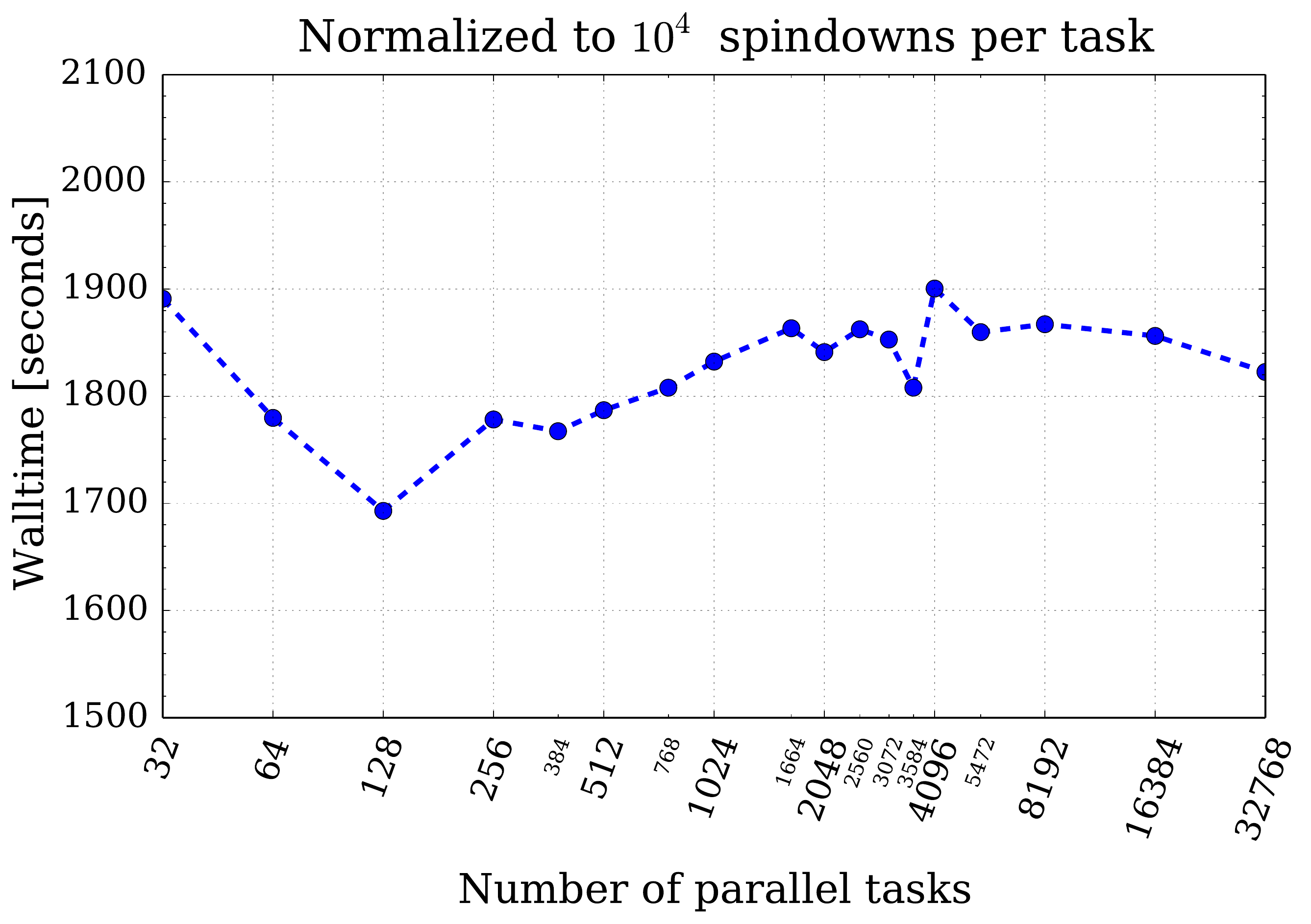}
	\caption{Weak scaling of the embedded \sky code. Each task involves 
	a search of $10^4$ spindowns. Average wall time to perform a task is shown 
	for an increasing number of parallel tasks.}
	\label{fig:weaksky}
\end{figure}

To determine how large a problem size could be we performed a weak scaling 
test by fixing the amount of work per processor - the ratio of spindown loop
computations to number of tasks - and compared the execution time. 
Fig.~\ref{fig:weaksky} shows the average time spent by a task consisting of
analysis of $10^4$ spindown loops as a function of the number of tasks on which
the computation is performed. This illustrates the fact that the 
combination of tasks at all scales, from 32 till 32768 tasks, into one 
parallel computation is efficiently realizable with time scales of around 
1900 seconds per frequency. 
By increasing the complexity of the problem by adding frequency bands to the 
same computation we were able to reach high scalability and test a run for 17 frequencies 
(from 100 till 2000) at once performed in only 56 minutes on 50272 tasks e.g., a 
half of the Cray Hermit supercomputer.

\subsection{Load balancing}
\begin{figure}
	\centering
	\includegraphics[width=\columnwidth]{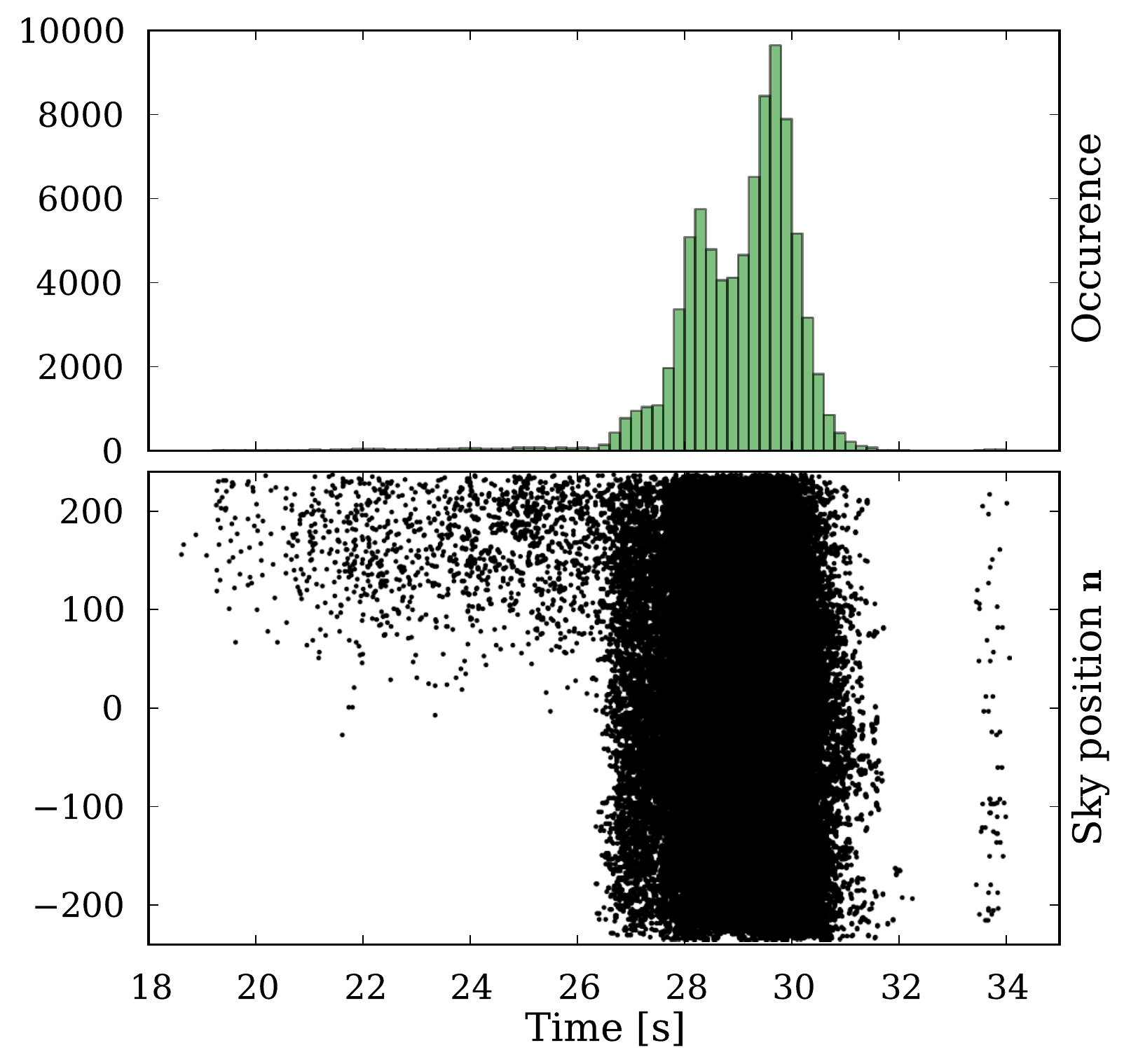}
	\caption{Relative distribution of computation for a search in each sky position, 
	showing an almost balanced computation, spending on average the same amount 
	of time per search. The unbalanced part is negligibly small.}
	\label{fig:loadbala}
\end{figure}

For an efficient use of the computational time it is essential that all the
cores finish their work as synchronously as possible. Locally on each
processor, this is done algorithmically since the coding and optimizing is made
to reach maximal speed-up by using the {\tt round-robin} algorithm. FFT computations
consisting of two loops over the sky positions, an inner loop 
over frequency derivative and finally the FFT that evaluates $\F$-statistic on a
grid of frequencies were distributed in parallel in a fully decoupled manner. 
Time scales for each parallel computation were approximately the same as it 
can be seen in Fig.~\ref{fig:loadbala}. 
The unbalanced part at some sky positions consist of computations that 
were about 30\% faster. But the number of these exceptionally
''light'' computations is less than 1\%. A possible non-optimal usage of the
computation power when some of the tasks spend more of their time in the waiting
state is decreasing with the number of computations per band and the latency 
is ''hidden'' by the scheduling system. By keeping the problem size big enough 
the sub-optimal usage of resources due to unbalanced computation is practically 
eliminated.

\section{Discussion and outlook}
\label{sect:summ}
We have parallelized the \sky code, developed by the Polgraw-Virgo group 
for the search of periodic GWs from spinning deformed neutron stars. 
The parallel version of the \sky embedded in \skyfarm presently scales 
up to many tens of thousands CPU cores allowing the GW search pipeline to run 
on massively parallel processing systems.

The farm skeleton system of \skyfarm is used for future improvement 
in developing massively scalable parallel versions for other scientific codes, 
e.g., parametric study of stability of nuclear clusters-isotopes at extreme
temperatures and density, and the code for simulation of ultra high energy cosmic rays \cite{HLRSworkshop}. 
The full benefit of using the non-collective communicator creation mechanisms 
presented by \skyfarm is worth further investigation when applying to other simulations 
and data analysis models. Reaching an extremely scalable computation \sky code 
would enable the data analysis for future Gravitational Wave experiments 
on acceptable time scales, 
using future supercomputing systems with millions of computing cores.

Parallelizing the \sky code allowed us to perform the search for gravitational waves 
at very high frequencies, for bands above 1200Hz. 
With a sequential code, analysis of our nominal 1Hz band at 1200Hz lasts more than one month.
Advanced parallelism of our skeleton system \skyfarm made it possible to do the analysis 
of many such high frequency bands at once in one massively parallel job.
  
The development of the \skyfarm was crucial to provide such a scalability. 
Scalability of our code is much higher comparing to the existing farm skeletons running on 
multi-node MPI systems. Here by the implementation of {\tt MPI\_Groups} and non-blocking 
global communications we solved a massive communication overhead problem. 
Using {\tt MPI\_Groups} for organizing a farm skeleton allowed us to incorporate 
the parallel version of the \sky code into a \skyfarm without a need of its global re-engineering. 

A potential ''bottleneck'' resulting from the unavoidable {\tt I/O} activity
rising with the number of nodes was neutralized for massive runs on over a
thousand cores by the implementation of a parallel MPI {\tt I/O}. This approach
however requires sufficiently fast communication between the nodes - the
optimal usage is limited to the HPC clusters with highly efficient
interconnects like InfiniBand.  In addition, we have re-engineered the
non-optimal memory management algorithm of the \sky code (it was initially
developed for sequential runs and was causing ''memory leaks'' when running as
an embedded part of the \skyfarm on massively parallel systems). 

In order to keep up with the huge size of the parameter space for analyzing the
future data of GW detector experiments, a hybrid parallelization of \skyfarm is
necessary. We anticipate that the resource-consuming part of the computation,
i.e., the FFT in \sky code, could be driven on the Graphical Processor Units
(GPU) or other hardware accelerators.
Hybrid parallelization of the \sky, as well as improved scheduling and
bookkeeping in the \skyfarm task farmer will also be crucial for
optimal use of the future massively parallel exascale computing systems that 
will be available in the next decade (facilities equipped with more than 100
million CPU cores and with multi-core architectures combining processors and 
co-processors).


\section{Acknowledgments}
\label{Acknowledgments}
The developments on parallelizing \sky, designing the \skyfarm, identifying the
bottlenecks and non-optimal usage of resources during simulations, were
possible by using massively parallel system Cray XE6 Hermit at High Performance
Computing Center HLRS in Stuttgart\footnote{Project ACID 12863}. The work of
G.~Poghosyan and A.~Streit were supported by the Helmholtz ''Supercomputing'' program. 
M.~Bejger and A.~Kr{\'o}lak thank the Steinbuch Centre for Computing of Karlsruhe Institute of
Technology for support and hospitability during their visits when this work was
done. The work of M.~Bejger and A.~Kr{\'o}lak was supported in part by the Polish Ministry of
Science and Higher Education grant DPN/N176/VIRGO/2009, and National Science
Center grants UMO-2013/01/ASPERA/ST9/00001 and 2012/07/B/ST9/04420. We thank
D.~Seldner for careful reading of this manuscript and useful comments.


\begin{thebibliography}{10}
\expandafter\ifx\csname url\endcsname\relax
  \def\url#1{\texttt{#1}}\fi
\expandafter\ifx\csname urlprefix\endcsname\relax\def\urlprefix{URL }\fi
\expandafter\ifx\csname href\endcsname\relax
  \def\href#1#2{#2} \def\path#1{#1}\fi

\bibitem{Ein1916}
A.~Einstein, In: Sitzungsberichte der K{\"o}niglich Preussischen Akademie der Wissenschaften Berlin (1916) 688.

\bibitem{HT75}
R.~A. {Hulse}, J.~H. {Taylor}, Astroph. J 195 (1975) L51.

\bibitem{Saul94}
P.~R. Saulson, Fundamentals of Interferometric Gravitational Wave Detectors, World Scientific, Singapore, 1994.

\bibitem{Astone2010}
P.~{Astone}, K.~M. {Borkowski}, P.~{Jaranowski}, M.~{Pi\k{e}tka}, A.~{Kr{\'o}lak}, \prd 82~(2) (2010) 022005.

\bibitem{Aasi:2014}
J.~{Aasi}, J.~{Abadie}, B.~P. {Abbott}, R.~{Abbott}, T.~D. {Abbott},  M.~{Abernathy}, T.~{Accadia}, et~al., LIGO Document No. LIGO-P1300133. \href{http://arxiv.org/abs/1402.4974} {\path{arXiv:1402.4974}}.

\bibitem{MPISpec3}
{MPI Forum}, A Message-Passing Interface Standard Version 3.0, High Performance Computing Center Stuttgart, 2012.

\bibitem{Aasi:2013a}
J.~{Aasi}, J.~{Abadie}, B.~P. {Abbott}, R.~{Abbott}, T.~D. {Abbott}, M.~{Abernathy}, T.~{Accadia}, et~al., LIGO Document No LIGO-P1200087. \href{http://arxiv.org/abs/1304.0670} {\path{arXiv:1304.0670}}.

\bibitem{JKS1998}
P.~{Jaranowski}, A.~{Kr{\'o}lak}, B.~F. {Schutz}, \prd 58~(6) (1998) 063001.

\bibitem{AasiEHS52013}
J.~{Aasi}, J.~{Abadie}, B.~P. {Abbott}, R.~{Abbott}, T.~D. {Abbott}, M.~{Abernathy}, T.~{Accadia}, et~al., \prd 87~(4) (2013) 042001.

\bibitem{RoundRobin}
L.~{Kleinrock}, Naval Research Logistics Quarterly 11~(1) (1964) 59-73.

\bibitem{dynmpi}
W.~Gropp, E.~Lusk, Seventh IEEE Symposium on Parallel and Distributed Processing (1995) 530--533.

\bibitem{colemit}
M.~Cole, Algorithmic Skeletons: Structured Management of Parallel Computation,  MIT Press, Cambridge, 1989.

\bibitem{Cole2004389}
M.~Cole, Parallel Computing 30~(3) (2004) 389 -- 406.

\bibitem{Ernsting2012129}
S.~Ernsting, H.~Kuchen, International Journal of High Performance Computing and Networking 7~(2) (2012) 129--138.

\bibitem{Gonzalez-Velez20101135}
H.~Gonzalez-Velez, M.~Leyton, Software - Practice and Experience 40~(12) (2010) 1135--1160.

\bibitem{dyngrp}
J.~Dinan, S.~Krishnamoorthy, P.~Balaji, J.~R. Hammond, M.~Krishnan, V.~Tipparaju, A.~Vishnu, Lecture Notes in Computer Science 6960 (2011) 282-291.
 
\bibitem{Rodgers}
D.~P. Rodgers, SIGARCH Comput. Archit. News 13~(3) (1985) 225--231.

\bibitem{HLRSworkshop}
G.~Poghosyan, S.~Sharma, A.~Kaur, V.~Jindal, P.~Bisht, et~al.,  in:  W.~E. Nagel, D.~B. Kröner, M.~M. Resch (Eds.), High Performance Computing in Science and Engineering '14, Springer International Publishing, 2014, pp. {\it in press ISBN 978-3-319-10809-4}.

\end{thebibliography}

\end{document}